\documentclass[11pt]{article}
\usepackage{graphicx}
\def\title{\begin{center}\Large\bf}
\def\author(s){\vspace{0.3cm}\large\rm}
\def\text{\end{center}}

\begin{document}


\noindent {\small{\it Submitted to the Proceedings of the
International Conference "Near-Earth Astronomy - 2007", 3-7
September 2007, Terskol settlement, Kabardino-Balkaria, Russia}}

\medskip \hrule \medskip

\bigskip
\bigskip


\title
Ukrainian Synchronous Network of small Internet Telescopes as rapid
action instrument for transient objects

\bigskip



\author(s)
B.E. Zhilyaev $^1$, M.V. Andreev $^2$, Ya.O. Romanyuk $^1$, A.V.
Sergeev $^2$, \\ V.K. Tarady $^2$

\bigskip

\smallskip

\noindent $^1$ {\small {\it Main Astronomical Observatory, NAS
of Ukraine,  27 Zabolotnoho, 03680 Kiev, Ukraine}} \\

\noindent {\small {\it e-mail:}} {\small{\bf zhilyaev@mao.kiev.ua}}

\smallskip

\noindent $^2${\small {\it International Centre for Astronomical,
Medical and Ecological Research \\ Terskol settlement,
Kabardino-Balkaria, 361605 Russia}} \\

\smallskip



\text


\section*{Abstract}

UNIT (The Ukrainian synchronous Network of small Internet
Telescopes) is a system of automated telescopes that search for
simultaneous optical activity of transient objects associated with
variable stars, small bodies of the Solar system, Near-Earth objects
(NEOs), gamma ray bursts, etc. Their instruments are sensitive down
to $M_{V} \approx 18$ and require an average of 60 seconds to obtain
the first images of the transient events after the alarm or GCN
notice. Telescopes of UNIT are equipped with fast CCD cameras to
study astrophysics on the timescales up to tens Hz. UNIT will be
operating by the middle of 2008.

\vspace{0.5 cm}

\noindent {{\bf keywords}$\,\,\,\,$\large instrumentation:
photometers -- instrumentation: detectors -- methods: observational
-- techniques: photometric -- telescopes }

\large

\section{Introduction}

The philosophy of UNIT is to develop an instrument allowing to
obtain observations through the Internet from a PC at any location.
UNIT is primarily supposed for professional applications. It can
also be employed for educational aims by students via a www gateway.
The foundation of this project must be the ideas of high technology,
innovative methods of observations and data processing, know-how.
Just such approaches allow the project will be able to survive in
the conditions of high competition in the world scientific
community.  A typical station of the Network is as follows: standard
equipment + trained operator + Internet. Correspondents and
participants of the Network can be organized in any country of the
world, including the Ukrainian station in Antarctic Continent.


\section{Instrumentation}

\noindent  UNIT consists of two observational complexes in Ukraine
and Russia (Peak Terskol, North Caucasus) complete by small
Celestron robotic telescopes with aperture 11 and 14 inches. We will
use frame-transfer CCDs as the detectors for UNIT. The cooled chips
have imaging areas of 1024x1024 and 512x512 pixels with resulting
dark current of 0.5 e-/pixel/s. For small windows it is possible to
take 0.01 s exposures. UNIT uses GPS technology to take the
pointing, guidance and observations in the synchronous operating
mode. Using GPS receiver, we will synchronize all exposures with two
remote UNIT telescopes to an absolute accuracy of better than 1
millisecond. We will use the UBVRI filter-set.


\section{Data acquisition}

\noindent The CCD software operates on Windows-based systems and
gives complete control over the image capture functions. The data is
transferred to a PC via a PCI bus (CCD47-10 AIMO Back Illuminated
Compact Pack High Performance CCD Sensor) and via an IEEE 1394
(FireWire) digital interface (the Rolera-MGi camera). The UNIT
telescopes run under local operator control. Observer communicates
with operator using the VoIP technology of real-time talk/call
transmissions through data networks. He conducts vocal or video
conversation, using connected to the computer microphone, loud
speaker and webcam. He can utilize the handwritten input for writing
of instantaneous reports through data network too.

\begin{figure*}
\centering
\resizebox{0.90\hsize}{!}{\includegraphics[angle=000]{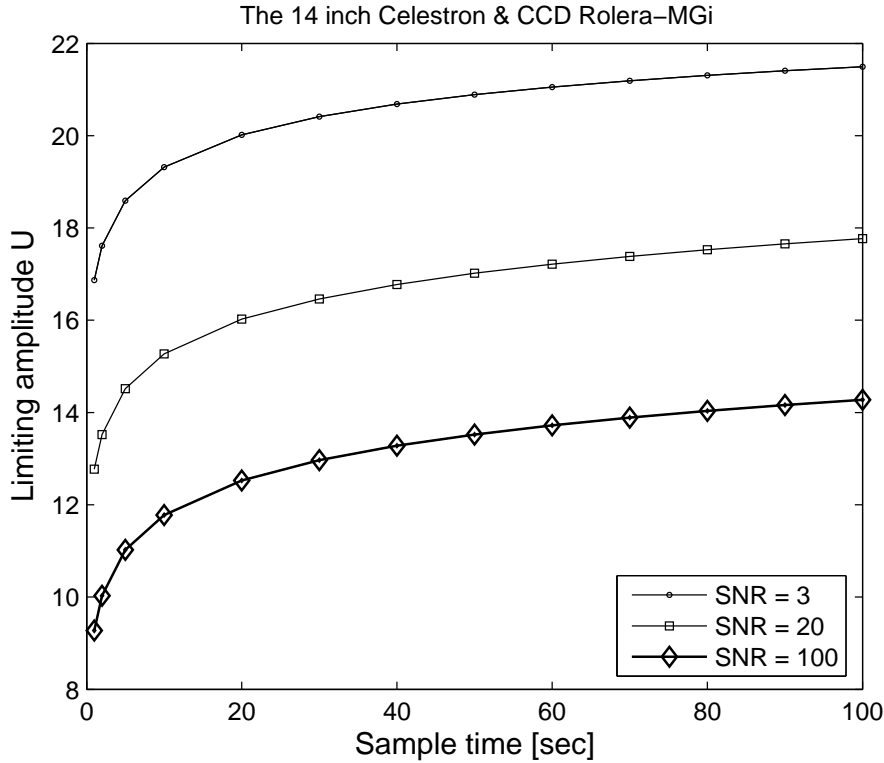}}%
\caption{Under average conditions, i.e. seeing of $\sim$ 1 arcsec, 2
min exposure a detectable magnitude is around U $\sim$ 21. It is
possible to measure stars up to 18 mag with the 5\% precision. Stars
of around 12 mag can be measured with the 1\% precision with 10 sec
exposure. Practical measurements at Peak Terskol (M. Andreev) fully
coincide with theoretical estimations.}
\end{figure*}

\begin{figure*}
\centering
\resizebox{0.90\hsize}{!}{\includegraphics[angle=000]{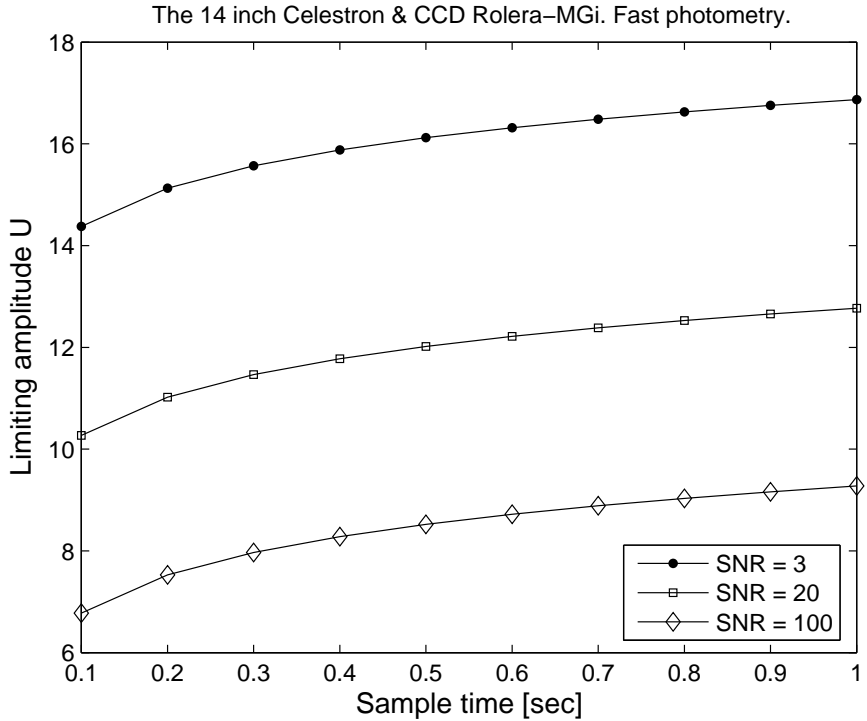}}%
\caption{Under seeing of $\sim$ 1 arcsec it is possible to carry out
fast stellar photometry within 5-50 Hz for stars of 10-13 magnitudes
with the precision of 10-20\%. Practical measurements with the 11
inch Celestron at Peak Terskol  proved that we could pick a
detectable magnitude 17.5 with 2 sec exposure. It is close to the
theoretical estimations.}
\end{figure*}


\section{Performance}

\noindent  Figure 1 shows the limiting U magnitudes for detection
with UNIT at a signal-to-noise of 3/20/100 as a function of exposure
time. Figure 2 shows the limiting U magnitudes for fast photometry
with the 14 inch Celestron.

\section{Science}

\noindent Foremost advances in astronomy can result when a new idea
becomes accessible for implementation. Distinctive feature of UNIT
consists in synchronous operation of several far remote telescopes
based on the robotic instrumentation. This allows studying wider
class of problems at other viewing angle. Studies of variables which
are 12 mags in quiescence on timescales of a second could be
accomplished with UNIT. We can use the coincidence technique with
the remote UNIT telescopes operated synchronously in order to obtain
the time resolution about 0.1 s, when studying transients, which are
typically 10-14 magnitudes. We can perform follow-up observations of
GRB optical-counterparts down to 18 magnitudes. The UNIT telescopes
are planned to be supplied with a low resolution optical
spectrograph (the resolving power $\sim 100$) for time-resolved
spectrography. This permits studying short-time variations on
timescales of seconds for stars of 10 magnitudes both in a
continuous spectrum and spectral lines simultaneously. UNIT has a
mission to work on around a dozen science programs:
\begin{enumerate}
  \item the activity of comets;
  \item searching for new asteroids, detection of NEOs, space
debris;
  \item wide range of observing programs including the fast
variability in light and spectra from the various types of variable
stars and centers of active galaxies, asteroseismology;
  \item the remotely operated long-term Stellar Flare
Monitor;
  \item follow-up observations of GRB
optical-counterparts;
  \item searching  for newly discovered Novae
and Supernovae, etc.
\end{enumerate}

\subsection{Some comparisons}

\begin{itemize}
  \item (1) TAROT are robotic observatories devoted to measure
  the early optical counterparts of gamma-ray bursts [1]. TAROT
  observe with no human interaction. TAROT are two identical
  25 cm telescopes $F/D=3.4$ that cover $1.86^{\circ}$ x $1.86^{\circ}$
  field of view on the Andor CCD cameras (Marconi 4240 back illuminated).
  Spatial sampling is 3.3 arcsec/pix. Six filters are available :
  BVRI, a clear filter and a 2.7 density coupled to V (for Moon and
  planets). Detection limit is about V=17 in 1 min. exposure.
  TAROT Calern observatory in France: $lon=6.9238^{\circ}$ E, $lat=43.7522^{\circ}$
  N, alt=1270 m. TAROT La Silla ESO observatory : $lon=70.7322^{\circ}$
  W, $lat=29.2608^{\circ}$ S, $alt=2347 m$.
  \item (2) Full automated Internet-telescope MASTER, Russia, Near Moscow,
  Alexander Krylov Observatory, Sternberg Astronomical Institute [2].
  Modified Richter-Slefogt Camera, D = 355 mm, D/F - 1:2.4,  Flat FOV -
  $5^{\circ}$ x $ 5^{\circ}$. CCD-camera AP16E (4000x4000). On optical observations
  of GRB: GRB040308 (GCN 2543) - 48 h after trigger time, OT limit 21.2 mag .
\end{itemize}

\subsection{Comparison with ULTRACAM}

\noindent ULTRACAM is an ultra fast camera capable of capturing some
of the most rapid astronomical events [3]. For comparison we employ
analogical results for the limiting magnitudes with ULTRACAM (at a
signal-to-noise of 10) as a function of exposure time at the GHRIL
focus of the 4.2-m William Herschel Telescope (WHT) on La Palma.
Difference between both our calculations and practical measurements
and of the ULTRACAM  data is explained only by a geometrical factor
- different diameters of telescopes - and equals about 4 magnitudes
(15 / 11 mag for the 0.1 s  time resolution and 13 / 9 mag for the
0.01 s one).

\bigskip
\noindent \large{\it Acknowlegement}

\bigskip
\noindent This work was supported  from the Science and Technology
Center in Ukraine (STCU), Project 4134.



\begin{thebibliography}{}

\small{ \baselineskip=1pt

\bibitem{Tarot}
TAROT. Telescopes a Action Rapide pour les Objets Transitoires,
http://tarot.obs-hp.fr/infos/

\bibitem{Master}
MASTER. Mobile Astronomical System of the TElescope-Robots,
http://observ.pereplet.ru/

\bibitem{ESO}
ESO Press Release 17/05, 9 June 2005

}

\end{thebibliography}
\end{document}